\documentclass{svmult}
\usepackage[latin9]{inputenc}
\usepackage{url}
\usepackage{graphicx}



\begin{document}

\title*{Beyond the quantum formalism: consequences of a neural-oscillator
model to quantum cognition}

\titlerunning{Beyond the quantum formalism}

\author{J. Acacio de Barros}

\institute{Liberal Studies Program, San Francisco State University, barros@sfsu.edu}

\maketitle

\abstract{In this paper we present a neural oscillator model of stimulus response
theory that exhibits quantum-like behavior. We then show that without
adding any additional assumptions, a quantum model constructed to
fit observable pairwise correlations has no predictive power over
the unknown triple moment, obtainable through the activation of multiple
oscillators. We compare this with the results obtained in reference
\cite{de_barros_decision_2013}, where a criteria of rationality gives
optimal ranges for the triple moment. }

\section*{Introduction}

Recently, much attention has been paid to quantum-mechanical formalisms
applied to human cognition (see \cite{busemeyer_quantum_2012,haven_quantum_2013,khrennikov_ubiquitous_2010},
and references therein). This comes from an increasing set of empirical
data better described by quantum models than classical probabilistic
ones (for an new effective classical approach, however, see \cite{dzhafarov_random_2013},
to appear in this proceedings). 

The underlying origins of such quantum-like features are not well
understood, but few researchers believe that actual quantum mechanical
processes are responsible (see \cite{busemeyer_quantum_2012} but
also \cite{pylkkanen_weak_2013} for a different view). Instead, as
argued in \cite{de_barros_quantum_2009}, what is behind such features
is a contextual influence. Interference-like effects in neuronal firings
in the brain lead to outcomes that are context dependent, similar
to the two-slit experiment in quantum mechanics, thus providing a
possible explanation. In fact, in \cite{de_barros_quantum-like_2012}
we showed how a simplified neural model with interference emerging
from the collective dynamics of coupled neurons gives origin to quantum-like
effects. Such model was designed to be consistent with currently known
neurophysiology and to reproduce the behavioral stimulus-response
theory \cite{suppes_phase-oscillator_2012}. Here, we discuss the
implications of such neural model to quantum cognition, and in particular
to the predictability power of the quantum-mechanical apparatus, as
opposed to its descriptive power.

\section*{Model and main results}

Here we briefly present the main model shown in \cite{de_barros_quantum-like_2012,suppes_phase-oscillator_2012},
and the readers are referred to them for details. For the simple case
of a continuum of responses, we start with representations of stimulus
and response in terms of phase oscillators. Such oscillators, made
out of collections of neurons, are synaptically coupled, and, depending
on the coupling strength, may synchronize. Let $s(t)$ be the neural
oscillator representing the activation of a stimulus, and $r_{1}(t)$
and $r_{2}(t)$ the oscillators for the two extremes in a continuum
or responses. We focus on their phases, $\varphi_{s}$, $\varphi_{r_{1}}$,
and $\varphi_{r_{1}}$, whose dynamics are given by 
\begin{eqnarray}
\dot{\varphi}_{i} & = & \omega_{i}+\sum_{j\neq i}k_{i,j}^{E}\sin(\varphi_{i}-\varphi_{j})+\sum_{j\neq i}k_{i,j}^{I}\cos(\varphi_{i}-\varphi_{j}),\label{eq:osc1}
\end{eqnarray}
where $k_{i,j}^{E}$ and $k_{i,j}^{I}$ are the overall excitatory
and inhibitory couplings between the neural oscillators. During reinforcement,
the coupling strengths $k_{i,j}^{E}$ and $k_{i,j}^{I}$ are changed
in a Hebb-like fashion. This model can easily be extended to include
multiple stimulus and response oscillators. For instance, in \cite{de_barros_quantum-like_2012}
it was used with two stimulus oscillators to obtain quantum-like effects.
Such effects were the consequence of couplings between the oscillators
that were reinforced to respond to two different stimuli corresponding
to incompatible contexts. When both stimuli were simultaneously activated,
an interference effect was obtained.

\begin{figure}
\sidecaption \includegraphics[width=7cm]{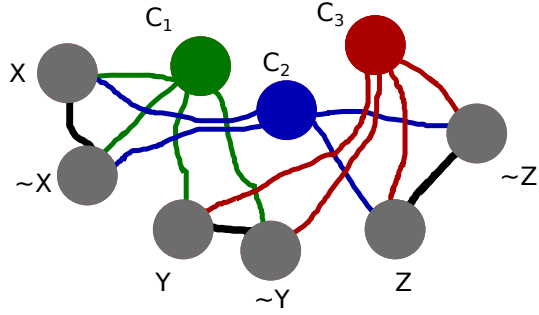}\protect\caption{\label{fig:Layout-of-oscillator}Layout of a neural-oscillator system
exhibiting pairwise correlations between $\mbox{\boldmath\ensuremath{X}}$,
$\mbox{\boldmath\ensuremath{Y}}$, and $\mbox{\boldmath\ensuremath{Z}}$.
In this oscillator system, $\left(\mathbf{X}=1\right)\&\left(\mathbf{Y}=-1\right)$
corresponds to the synchronization with oscillator $C_{1}$ closer
in phase to $X$ and not to $\sim X$, while at the same time being
closer to $Y$ than to $\sim Y$. }
\end{figure}

Quantum-like models lead to contextual responses, in the sense that
there exists no joint probability distribution for the associated
random variables. Let us look at the particular example presented
in reference \cite{de_barros_joint_2012} and expanded in another
context in \cite{de_barros_decision_2013}. Let $\mbox{\boldmath\ensuremath{X}}$,
$\mbox{\boldmath\ensuremath{Y}}$, and $\mbox{\boldmath\ensuremath{Z}}$
be $\pm1$-valued random variables, and consider the neural oscillator
system represented in Figure \ref{fig:Layout-of-oscillator}. 
 For this system, the activation of one of the three stimulus oscillators,
$C_{1}$, $C_{2}$, or $C_{3}$, leads to the corresponding responses
computed via phase differences. For example, if $C_{1}$ is sampled,
the oscillators' dynamics, dictated by the specific values of inhibitory
and excitatory couplings, converge to a fixed point that may favor
$\mathbf{X}=1$ (oscillator $X$) instead of $\mathbf{X}=-1$ (oscillator
$\sim X$), while at the same time favoring $\mathbf{Y}=-1$, thus
corresponding to a negative correlation. With such oscillator system,
it is possible in principle to choose couplings such that the correlations
between $\mbox{\boldmath\ensuremath{X}}$, $\mbox{\boldmath\ensuremath{Y}}$,
and $\mbox{\boldmath\ensuremath{Z}}$ are too strong for a joint probability
distribution to exist. As a consequence, and because of the pairwise
commutativity of the set of quantum-mechanical observables $\hat{X}$,
$\hat{Y}$, and $\hat{Z}$ corresponding to the random variables $\mbox{\boldmath\ensuremath{X}}$,
$\mbox{\boldmath\ensuremath{Y}}$, and $\mbox{\boldmath\ensuremath{Z}}$,
it follows that there exists no state $|\psi\rangle$ in the Hilbert
space ${\cal H}$ where such observables are defined and such that
the neural correlations hold \cite{de_barros_decision_2013}. However,
even in such situations a quantum model can be constructed \cite{aerts_quantum_2013},
and in order to describe the correlations set by the neural-oscillator
model, we are forced to expand the Hilbert space to ${\cal H}'\otimes{\cal H}$
\cite{de_barros_decision_2013}. For instance, we can write a state
vector 
\begin{equation}
|\psi\rangle=c_{xy}|A\rangle|\psi_{xy}\rangle+c_{xz}|B\rangle|\psi_{xz}\rangle+c_{yz}|C\rangle|\psi_{yz}\rangle,\label{eq:quantum-state}
\end{equation}
where $|A\rangle$, $|B\rangle$, and $|C\rangle$ are orthonormal
vectors in ${\cal H}'$, $\langle\psi_{xy}|\hat{X}\hat{Y}|\psi_{xy}\rangle=-2/3$,
$\langle\psi_{xz}|\hat{X}\hat{Z}|\psi_{xz}\rangle=-1/2$, $\langle\psi_{yz}|\hat{Y}\hat{Z}|\psi_{yz}\rangle=0$,
and $c_{xy}$, $c_{xz}$, and $c_{yz}$ are such that $\left|c_{xy}\right|^{2}+\left|c_{xz}\right|^{2}+\left|c_{yz}\right|^{2}=1$.
Because each of the states $|\psi_{xy}\rangle$, $|\psi_{xz}\rangle$,
and $|\psi_{yz}\rangle$ can have arbitrary triple moments (they do
not fix enough of the distribution) between $-1$ and $1$, it follows
that (\ref{eq:quantum-state}) can describe the correlations but has
no predictive power with respect to the neural oscillator model or
human decision making.

However, the couplings encoding different responses in the Kuramoto
equations do determine, within a certain range, values for the triple
moment. The triple moment would be the equivalent, following \cite{de_barros_quantum-like_2012},
of a simultaneous activation of all stimulus oscillators. Thus, the
neural model would provide a definite prediction, in contrast with
the quantum one.

\section*{Final remarks}

Quantum formalisms applied to human cognition have shown a great potential
for certain applications in the social sciences. However, one must
ask how this is so, and also how predictive they are. For instance,
as showed above, it is possible to devise an neural system whose quantum
description has no predictive power. Thus, we could in principle design
an experiment to test this neural system, but not its corresponding
quantum description. 

Could be some principle to be added to the quantum description that
could provide predictions for outcomes of the experiment proposed?
For example, in \cite{de_barros_decision_2013} we proposed a minimization
principle as a normative decision for quantum-like inconsistencies,
which allowed signed probabilities to move from a descriptive to a
normative theory. Perhaps a principle of this type added to the quantum
formalism could be not only normative but predictive as well. However,
if we think that the underlying dynamics for quantum cognition is
actually from the complex and contextual interaction of neurons, perhaps
some similar principle from it should be added to the quantum description. 

Finally, we would like to emphasize that the quantum approach has
suggested interesting experiments in psychology. As such, it is a
promising field not only because of its ability to describe experiments,
but also for the intuitions it provides for thinking about context-rich
situations. Therefore, understanding its limitations and perhaps extending
it would be desirable. 
\begin{acknowledgement}
We thank Sandro Sozzo for pointing out reference \cite{aerts_quantum_2013},
and him as well as Gary Oas, Ehtibar Dzhafarov, Paavo Pilkkanen, and
Sisir Roy for useful comments and discussions. 
\end{acknowledgement}
\bibliographystyle{spmpsci}
\bibliography{QuantumBrain}

\end{document}